\documentclass[fleqn,10pt]{wlscirep}
\usepackage{amssymb}
\usepackage{graphicx}
\usepackage{dcolumn}
\usepackage{bm}
\usepackage{color}

\begin{document}

\newcommand{\ba}{{\bf a}}
\newcommand{\BB}{{\bf b}}
\newcommand{\bd}{{\bf d}}
\newcommand{\br}{{\bf r}}
\newcommand{\bp}{{\bf p}}
\newcommand{\bk}{{\bf k}}
\newcommand{\bg}{{\bf g}}
\newcommand{\bt}{{\bf t}}
\newcommand{\bu}{{\bf u}}
\newcommand{\bq}{{\bf q}}
\newcommand{\bG}{{\bf G}}
\newcommand{\bP}{{\bf P}}
\newcommand{\bJ}{{\bf J}}
\newcommand{\bK}{{\bf K}}
\newcommand{\bL}{{\bf L}}
\newcommand{\bR}{{\bf R}}
\newcommand{\bS}{{\bf S}}
\newcommand{\bT}{{\bf T}}
\newcommand{\bQ}{{\bf Q}}
\newcommand{\bA}{{\bf A}}
\newcommand{\bH}{{\bf H}}
\newcommand{\bX}{{\bf X}}

\newcommand{\bra}[1]{\left\langle #1 \right |}
\newcommand{\ket}[1]{\left| #1 \right\rangle}
\newcommand{\braket}[2]{\left\langle #1 | #2 \right\rangle}
\newcommand{\mel}[3]{\left\langle #1 \left| #2 \right| #3 \right\rangle}

\newcommand{\bdel}{\boldsymbol{\delta}}
\newcommand{\bsig}{\boldsymbol{\sigma}}
\newcommand{\beps}{\boldsymbol{\epsilon}}
\newcommand{\bnu}{\boldsymbol{\nu}}
\newcommand{\bnab}{\boldsymbol{\nabla}}
\newcommand{\bchi}{\boldsymbol{\chi}}
\newcommand{\bGam}{\boldsymbol{\Gamma}}

\newcommand{\bgt}{\tilde{\bf g}}

\newcommand{\brh}{\hat{\bf r}}
\newcommand{\bph}{\hat{\bf p}}


\title{Anomalous Dirac point transport due to extended defects in bilayer graphene}

\author[1,*]{Sam Shallcross}
\author[2]{Sangeeta Sharma}
\author[3,4]{Heiko B. Weber}
\affil[1]{Lehrstuhl f\"ur Theoretische Festk\"orperphysik, Staudstr. 7-B2, 91058 Erlangen, Germany,}
\affil[2]{Max-Planck-Institut f\"ur Mikrostrukturphysik Weinberg 2, D-06120 Halle, Germany,}
\affil[3]{Lehrstuhl f\"ur Angewandte Physik, Staudtstr.\ 7,
91058 Erlangen, Germany.}
\affil[4]{Interdisziplin\"ares Zentrum f\"ur Molekulare Materialien, Friedrich-Alexander-Universit\"at Erlangen-N\"urnberg}
\affil[*]{sam.shallcross@fau.de}

\date{\today}


\begin{abstract}

Charge transport at the Dirac point in bilayer graphene exhibits two dramatically different transport states, insulating and metallic, that occur in apparently otherwise indistinguishable experimental samples. We demonstrate that the existence of these two transport states has its origin in an interplay between evanescent modes, that dominate charge transport near the Dirac point, and disordered configurations of extended defects in the form of partial dislocations. In a large ensemble of bilayer systems with randomly positioned partial dislocations, the conductivity distribution $P(\sigma)$ is found to be strongly peaked at both the insulating and metallic limits. We argue that this distribution form, that occurs only at the Dirac point, lies at the heart of the observation of both metallic and insulating states in bilayer graphene.

\end{abstract}

\maketitle


\section{Introduction}

Transport at the Dirac point in structurally perfect bilayer graphene is expected to exhibit a minimal metallic conductivity, very similar in nature to that found in single layer graphene~\cite{Twor,Snyman,McCann}. Sharply contradicting this expectation, however, recent transport experiments on ultra-clean suspended bilayer graphene report instead the existence, near charge neutrality, of an insulating phase. Curiously, this typically occurs in only about 50\% of a nominally identical set of high mobility samples, with the remainder showing the expected minimal metallic conductivity~\cite{Bao,Feldman,Freitag,Mayorov,San-Jose}. Two different explanations for the existence of the insulating state have been proposed: (i) that the 8-fold degeneracy of the bilayer electron fluid at the Dirac point is responsible for an interaction driven symmetry breaking to an insulating phase~\cite{Bao,Feldman,Freitag,Mayorov} or, alternatively, (ii) that the insulating state is a manifestation of charge blocking by recently imaged\cite{Alden,Butz,Kisslinger,Ju} partial dislocations~\cite{San-Jose}.

Neither of these theories, however, has directly addressed what is perhaps the most remarkable aspect to emerge from experiment: the existence of dramatically different transport states - from metallic to insulating behaviour - in samples that are, so far as can be characterized, essentially identical. This is the issue that will form the focus of this work. Our key finding is that in an ensemble of systems with randomly placed partials there exists a pronounced tendency towards either strongly insulating states or minimally metallic states - i.e., some random configurations of partials entirely block transport while others appear not to impede charge transport at all. This behaviour, which sharply deviates from the well known paradigm that disorder should generally act to suppress transport, we argue underpins the observation of two transport states.

\section{Model}

\begin{figure}[t!]
    \centering
    \includegraphics[width=0.9\textwidth]{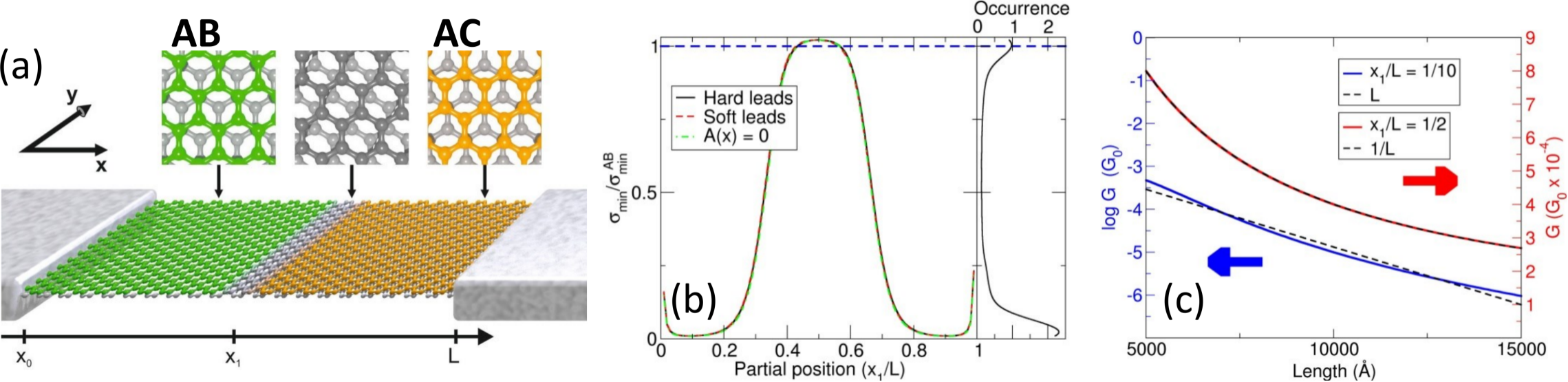}
    \caption{\emph{Resonant and ``blocked'' transport states for a single partial dislocation in bilayer graphene}: (a) Schematic illustration of a bilayer graphene ribbon consisting of two domains of structurally perfect Bernal stacking connected by a single partial dislocation; upper panels display the structure of the local AB and AC stacking geometries, while the grey area depicts the finite non-Bernal transition region at the partial. (b) Conductivity $\sigma$ at the Dirac point as a function of the partial position $x_1/L$, ranging from minimal metallic conductivity at $x_1/L = 1/2$ to insulating behavior at $x_1/L = 2/10$. Calculations are shown both for ``hard wall'' leads\cite{Snyman} as well as soft leads in which the transition from lead to sample has an exponential decay envelope, and also for the case in which a strain field induced by the partial is switched off ($A(x) = 0$). The right hand plot shows the likelihood of occurrence for a given transport state. (c) Dependence of the conductance $G$ on sample length for the metallic ($x_1/L = 1/2$) and insulating ($x_1/L=2/10$) partial positions.}
    \label{1}
\end{figure}

The microscopic origin of partial dislocations arises from the fact that Bernal stacking may be achieved in two equivalent ways, typically referred to as AB and AC stacking, as shown in Fig.~\ref{1}a. While evidently equivalent in a bilayer of infinite extent, their difference is physically significant if domains of both types coexist in the same sample. Continuity of the lattice requires that domains of different stacking order are connected by \emph{partial dislocations}, localized regions in space in which the transition between the different stacking types of the distinct domains is made. Three such partial dislocations are found in bilayer graphene\cite{Butz,Alden,Kisslinger,Ju}, described by one of the partial Burgers vectors: $\bd_1 = a \left[1/2,1/(2\sqrt{3})\right]$, $\bd_2 = a \left[0,-1/\sqrt{3}\right]$, and $\bd_3 = a \left[-1/2,1/(2\sqrt{3})\right]$ (where $a$ is the graphene lattice parameter, and $\bd_i$ the nearest neighbour vectors of the honeycomb lattice).

In order to understand the impact of such a domain structure on charge transport, we will consider a model of partial dislocations that are parallel with the left and right electrodes contacting the bilayer sample (see Fig.~\ref{1}a). This is the simplest conceivable structural model, and has advantages both in tractability (as it results in an effectively one dimensional model) as well as, most importantly for our purpose here, lending itself to a transparent analysis of the role of structural disorder. We note that while the imposition of a straight partial geometry might appear a rather restrictive condition, substantially simplifying mode matching at the partials, the difference between straight and non-straight partials was found not to be qualitatively significant in Ref.~\citeonline{San-Jose}. We will return to this point subsequently. Even with this simplifying assumption, however, modeling the transport experiments remains a challenge due to the presence of two very different length scales: the partial dislocation which occurs on a scale of $\approx$~5nm, and the dimensions of the suspended bilayer graphene sample, which has a length typically of the order of 1$\mu$m. This is therefore a multi-scale problem, and so evidently precludes the use of an atomistic approach, such as the tight-binding method. Instead, we will make use of the use of a recently developed effective Hamiltonian theory designed to treat structural deformations in low dimensional systems\cite{M}, and shown to provide a very good description of the electronic structure both for partial dislocations\cite{Kisslinger,M} as well as twist faults in bilayer graphene\cite{shall10,shall16}. In this scheme the individual layers of the bilayer are modeled by the Dirac-Weyl Hamiltonian $H_0 = v_F \bsig.\bp$ (with gauge fields arising from the strain fields of the partial included), while the spatially dependent stacking order is represented by an ``interlayer stacking field'' that, importantly, treats all stacking types on an equal footing. Further details may be found in the Computation Details section.

\section{Transport through a single partial dislocation}

\begin{figure}[t!]
    \centering
    \includegraphics[width=0.75\textwidth]{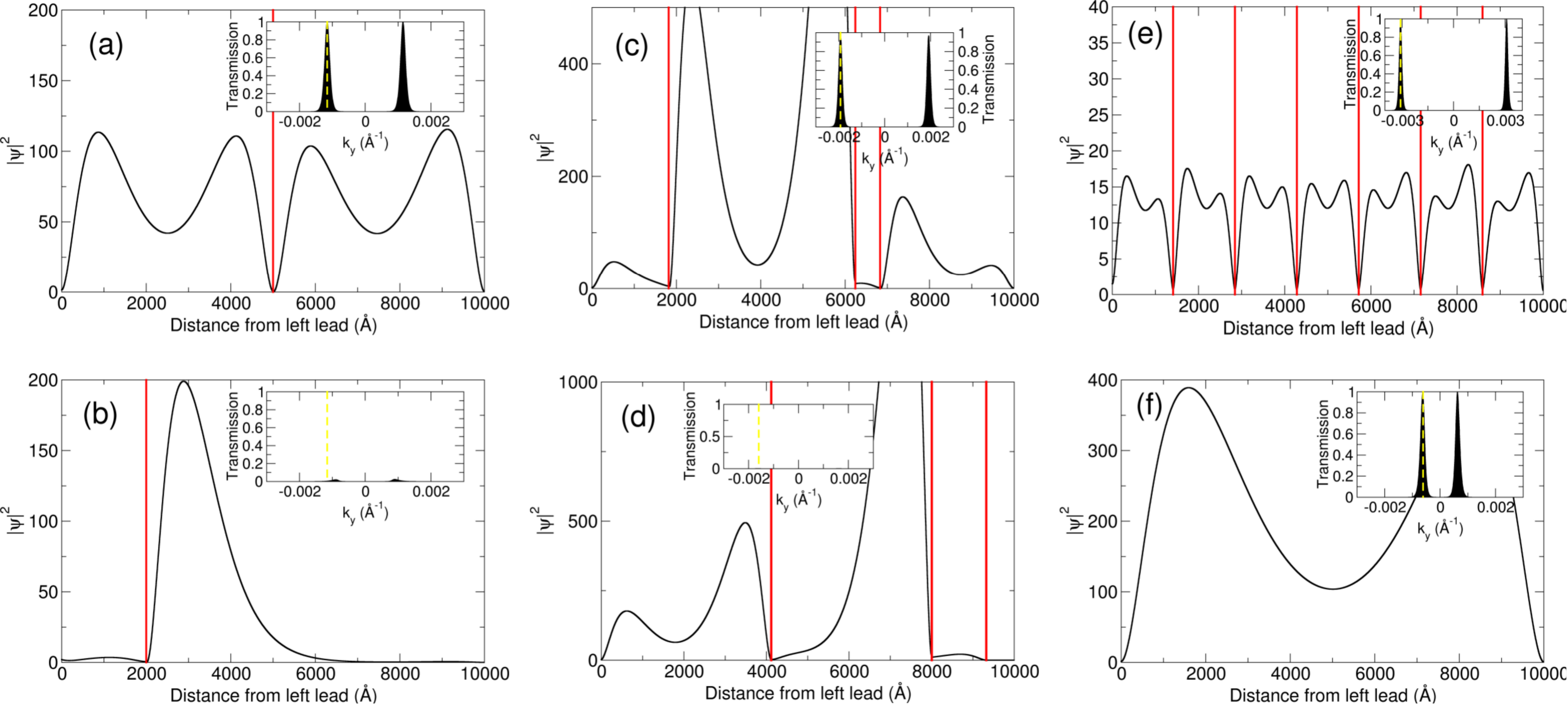}
    \caption{Transport states $|\Psi(x)|^2$ at the Dirac point for six partial dislocation configurations in a 1$\mu$m sample of bilayer graphene. The vertical red lines represent the positions of the partial dislocations, which in each case can be seen to determine the basic structure of the wavefunction. Panels (a) and (b) show, respectively, minimal conductivity and insulating states for a single partial, panels (c) and (d) minimal conductivity and insulating states for a three partial geometry, panel (e) the transport state for an ordered configuration of 6 partials, and lastly panel (f) for the pristine bilayer (zero partials). The inset panels display the transmission function $T$ for each partial configuration as a function of lead momentum $k_y$, with the lead momenta for which the transport state is plotted indicated by the vertical yellow line.}
    \label{wf}
\end{figure}

We first apply this scheme to the single partial configuration of Fig.~\ref{1}a, finding a large range of minimal conductivities as a function of partial position (Fig.~\ref{1}b). These include both a slightly resonantly enhanced conductivity when the partial is at the high symmetry central position, as well as an essentially insulating behaviour when the partial moves out of this region towards one of the leads. To confirm the metallic and insulating character of these two transport states we examine the dependence of the conductance $G$ on the system length finding, as expected, for the insulating case ($x_1/L=2/10$) an exponential decay with length, and for the metallic case ($x_1/L = 1/2$) an algebraic $L^{-1}$ decay with length (see Fig.~\ref{1}c). The resonant metallic state is, in fact, a simple generalization of that found for the structurally perfect AB bilayer by Snyman \emph{et al.}, and indeed the Fano factor for the resonant state is one third, just the value found for both the structurally perfect bilayer and single layer graphene. Strikingly, this simple model already shows the preferential occurrence of two transport states, as may be seen from the probability of occurrence $P(\sigma)$ (see Fig.~\ref{1}b). Integrating $P(\sigma)$ over a conductivity interval yields the probability that a randomly placed partial would result in a conductivity in that range and, as may be seen, this function has pronounced peaks at both the insulating and minimally metallic transport states. This physics, it should be stressed, is rather independent of both the lead structure and the strain state of the partial (as may be seen in Fig.~\ref{1}b). Furthermore, trigonal warping has no significant effect on the transport  due to the very small momenta of the satellite Dirac points.

To understand the physics behind these two transport states we examine the form of the corresponding sample wavefunctions (panels (a) and (b) of Fig.~\ref{wf}). The occurrence of minimal metallic conductivity is seen to be  associated with an evanescent resonance, of a very similar nature to that which occurs in a sample with no partial dislocations (panel (f) of Fig.~\ref{wf}). This resonance, once the partial dislocation is moved to a low symmetry position, is suppressed thus blocking the transport. In short, the existence of two transport states in bilayer graphene would appear, for the case of a single straight partial, to arise from the usual "particle in a box" physics of evanescent resonant states. Once more than one partial is present the realization of a high symmetry structure analogous to the central partial position is highly unlikely, and one might therefore expect that disordered arrangements of partials would, in all cases, lead to transport blocking. Interestingly, as we now show, this is not the case.

\section{Transport in the case of multiple partial dislocations}

\begin{figure}[t!]
    \centering
    \includegraphics[width=0.94\textwidth]{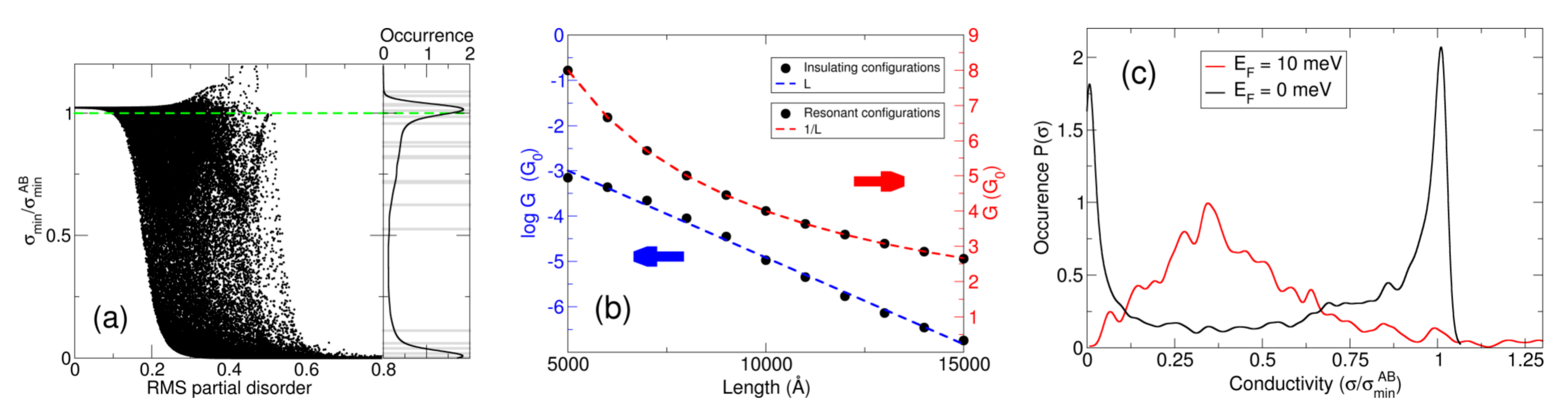}
    \caption{\emph{Insulating and metallic behaviour in evanescent transport through disordered configurations of partial dislocations.} (a) For three partials, the minimal conductivity at the Dirac point is computed for an ensemble of randomly chosen partial positions, plotted versus the RMS of the domain length distribution $L_{rms}^2 = \sum_{i=1,4} (L_i-\overline{L})^2/\overline{L}^2$ of the partial configuration. Each dot represents one of 76,210 calculated realizations of the positions of three partial dislocations in a 1$\mu m$ sample. All transport states between resonant and fully insulating may be seen, however minimal metallic conductivity (indicated by the dashed green line) and fully insulating states are significantly more likely to be found than any other, as shown in the probability of occurrence function in the right hand panel. The grey bars represent experimental data\cite{Bao}, which show a qualitatively similar likelihood of occurrence of transport states (each bar represents a transport measurement in one suspended bilayer graphene sample). (b) Conductance as a function of sample length: for a restricted ensemble average over configurations close to minimal metallic conductivity (defined by $|\sigma_{min}/\sigma_{min}^{AB} -1| < 10^{-3}$) a metallic $L^{-1}$ dependence on length is seen, while for a restricted ensemble average performed over the insulating configurations ($|\sigma_{min}/\sigma_{min}^{AB}| < 10^{-3}$) the expected exponential decrease in conductivity with $L$ is found ($\sigma_{min}^{AB}$ represents the minimal metallic conductivity of bilayer graphene). In panel (c) is shown the probability of occurrence function at the Dirac point and at a finite doping of 10~meV -- the striking difference in forms may immediately be observed.}
    \label{3}
\end{figure}

For more than one partial, the generic case is disorder amongst the partial coordinates $\{x_i\}$ which, as each partial is defined by one coordinate, may be characterized by a single number - the RMS of the terrace length distribution  $L_{rms}^2 = \sum_i (L_i-\overline{L})^2/\overline{L}^2$. In Fig.~\ref{3}(a) we consider the case of 3 partial dislocations, and plot an ensemble of 76,210 randomly generated configurations against this quantity. Three distinct regions of transport versus $L_{rms}$ may be seen. For small $L_{rms} < 0.1$, corresponding to ordered configurations in which the partials are approximately uniformly spaced, the transport resembles that of the high symmetry single partial configuration: a few percent resonant enhancement over the minimal metallic conductivity of pristine bilayer graphene. This is the multi-partial geometry directly analogous to the central position in the case of a single partial. Contrasting this, in the limit of large $L_{rms} > 0.65$ the transport is seen to be completely blocked. This $L_{rms}$ regime corresponds to case of a "partial pairing" geometry in which two of the three partials have come close to their closest allowed separation of $125\,$\AA. Most interesting, however, is the intermediate no-pairing but disordered regime. In this region the same value of $L_{rms}$ is found to yield almost all transport states from quite strong resonant enhancement ($\approx 20\%$ above the conductivity of structurally perfect bilayer graphene) to complete suppression of transport. The probability of occurrence function $P(\sigma)$ (Fig.~\ref{3}a) once again shows pronounced peaks at both minimal metallic conductivity and strong suppression of transport, just as in the case of a single partial dislocation. Two state transport is thus robust against disorder amongst partial coordinates for multi-partial configurations -- a very surprising result if one considers that in the case of a single partial the two state transport was driven by a high symmetry resonant state. This two state transport, it should be stressed, is a feature of transport at the Dirac point, and does not hold at any finite energy (see Fig.~\ref{3}c for a comparison of $P(\sigma)$ at the Dirac point and at a finite energy). Finally, to make a direct comparison with experiment we take the data from Ref.~\citeonline{Bao} which, after rescaling, is presented by the grey bars in the right hand panel of Fig.~\ref{3}a. Each bar represents a transport measurement in a single suspended bilayer graphene sample, and the experimental distribution of conductivities is seen to match rather well the theoretical distribution function.

This behaviour is both surprising in view of the resonant origin of two state transport in the single partial case, as well as markedly different from that which would be expected in the case of oscillatory transport states with finite momentum, for which the presence of disorder would generically result in destructive interference and the suppression of transport, not two state transport. To probe the reason behind this difference we plot the system wavefunction $\Psi(x)$ for several partial configurations (Figs.~\ref{wf}a-f), with the transport state indicated by the transmission plot in each case (see insets to Figs.~\ref{wf}a-f). Strikingly, in all cases the wavefunction takes its basic form entirely from the geometry of the partial configuration: each AB or AC domain has a similar $|\Psi(x)|^2$ structure, which differs only in magnitude between different terraces. This reflects the fact that evanescent states are waveforms without a phase structure and, as a consequence, the usual paradigm of destructive interference does not hold. As a result of this, and as may be seen in Fig.~\ref{wf}c,d for example, disordered configurations of partials may both support resonant states as well as leading to transport blocking. It is this that underpins the occurrence of two transport states in an ensemble of multi-partial configurations.

To further connect with experiment we now investigate the physical properties of the insulating configurations seen in Fig.~\ref{3}(a). To this end we first define, from the full ensemble, an insulating ensemble consisting only of states that satisfy $|\sigma_{min}/\sigma_{min}^{AB}| < 10^{-3}$. We firstly confirm the insulating character of this reduced ensemble from the exponential form of the ensemble averaged $\left<\sigma(L)\right>$ (see Fig.~\ref{3}(b)). Similarly, we may define ``minimal metallic ensemble'' by the restriction $|\sigma_{min}/\sigma_{min}^{AB} -1| < 10^{-3}$ which, reassuringly, displays the expected algebraic $L^{-1}$ behaviour characteristic of a metal (see Fig.~\ref{3}(b)). We now consider the impact on the insulating ensemble of (a) introducing a layer perpendicular field $E_\perp$ and (b) changing the particle number on members of the insulating ensemble.

\begin{figure}
    \centering
    \includegraphics[width=0.7\textwidth]{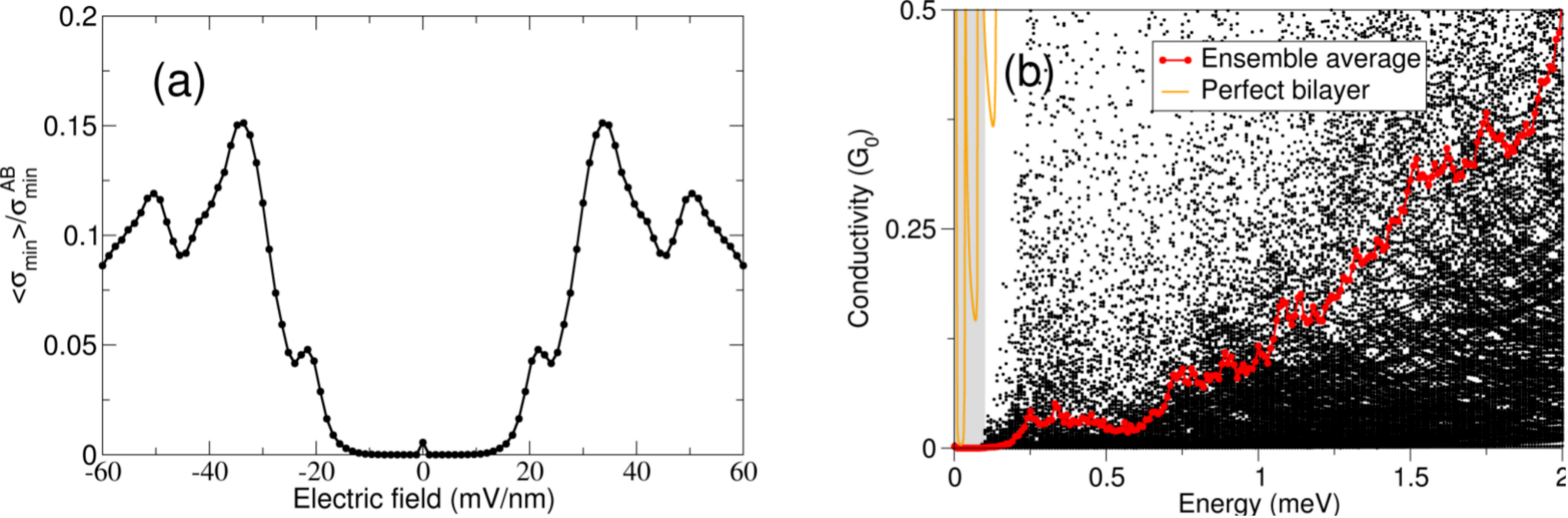}
    \caption{Dependence of the partial dislocation induced insulating state on (a) layer symmetry breaking potential $V$ and (b) the Fermi energy $E_F$. In both panels the conductivities are obtained by averaging over a large ensemble of partial configurations that, for zero field and at charge neutrality, result in an insulating state (defined by $\sigma_{min}/\sigma_{min}^{AB} < 0.001$, where $\sigma_{min}^{AB}$ represents the minimal metallic conductivity of bilayer graphene). In (b) the individual points represent members the ensemble of conductivities, while the red line denotes the ensemble average. The grey shaded region indicates the energy window within which transport is blocked for all partial configurations.}
    \label{ve}
\end{figure}

Application of a layer perpendicular field $E_\perp$ weakly restores the metallic state, see Fig.~\ref{ve}a, which in our model follows simply from the fact that an applied bias results in the well known ``Mexican hat'' dispersion~\cite{McCann}, thus imbuing the AB/AC segments with a small but finite momentum and so opening a propagating channel between them. As in experiment~\cite{Weitz,Velasco}, further increase in $E_\perp$ results once again in a suppression of the minimal metallic state.  Increase in the bilayer density by top gating is also known to restore transport and, as may be seen in Fig.~\ref{ve}b, our model also captures this effect with, most interestingly, a transport gap -- within which the insulating ensemble remains fully insulating for all members -- of a similar magnitude (0.2~meV) to that reported in experiment~\cite{Freitag,Bao,Feldman}. This has a qualitatively similar origin as restoration by field: an increase in energy will allow real momenta in the transport direction (as one is no longer at the Dirac point) and so will reopen propagating channels in the transport.

In both Figs.~\ref{ve}a and \ref{ve}b it will be noted that the reestablishment of the metallic state is rather weak, a fact which prompts a discussion of the straight partial model we have employed. The inherent translational symmetry of this model results in a separation of real lead momenta ($k_y$) from, at the Dirac point, imaginary transport momenta ($i k_x$), which in an irregular partial geometry will mix. At the Dirac point, however, states of real momenta are blocked by partials\cite{San-Jose} and so we therefore expect our Dirac point results to be robust to this more complex mode matching. On the other hand at finite doping the re-established propagating channel is expected to generate a higher conductivity in an irregular partial geometry than in a straight partial geometry, precisely due to this mixing of momenta. We thus expect the weaker establishment of the metallic state at finite doping and finite symmetry breaking field to be the most significant impact of the straight partial approximation.

\section{Summary and discussion}

We have shown that the existence of both conducting and insulating transport states in apparently indistinguishable experimental samples of bilayer graphene finds a natural explanation in terms of structural imperfections in the form of partial dislocations and transport by evanescent states. The absence of a phase structure in such states results in an unusual behaviour in which both ordered as well as disordered configurations of partial dislocations may support evanescent resonances, a behaviour that is, needless to say, very different to the usual physics of resonance. These Dirac point resonant states result in a minimal metallic conductivity very close to that of the pristine bilayer and, in their absence, transport is blocked. Over an ensemble of possible partial configurations one therefore finds a pronounced tenancy to either minimal metallic conductivity or very strong suppression of transport. The dramatically different transport states found in experiment are thus simply a manifestation of this ``hidden'' structural degree of freedom, which will naturally change from sample to sample as well as be changed by annealing the same sample. Finally, we note that ultra-clean suspended bilayer graphene has, in addition to interesting zero field transport physics, interesting physics at finite magnetic fields, in particular possible indications of a fractional quantum Hall effect\cite{Kou,Maher,LeRoy}. In the light of the convincing role that structural disorder plays in explaining the physics at zero field, it will be of interest to examine the impact of structural disorder at finite fields.

\section{Computational details}

\begin{figure}[thbp]
    \centering
    \includegraphics[width=0.5\textwidth]{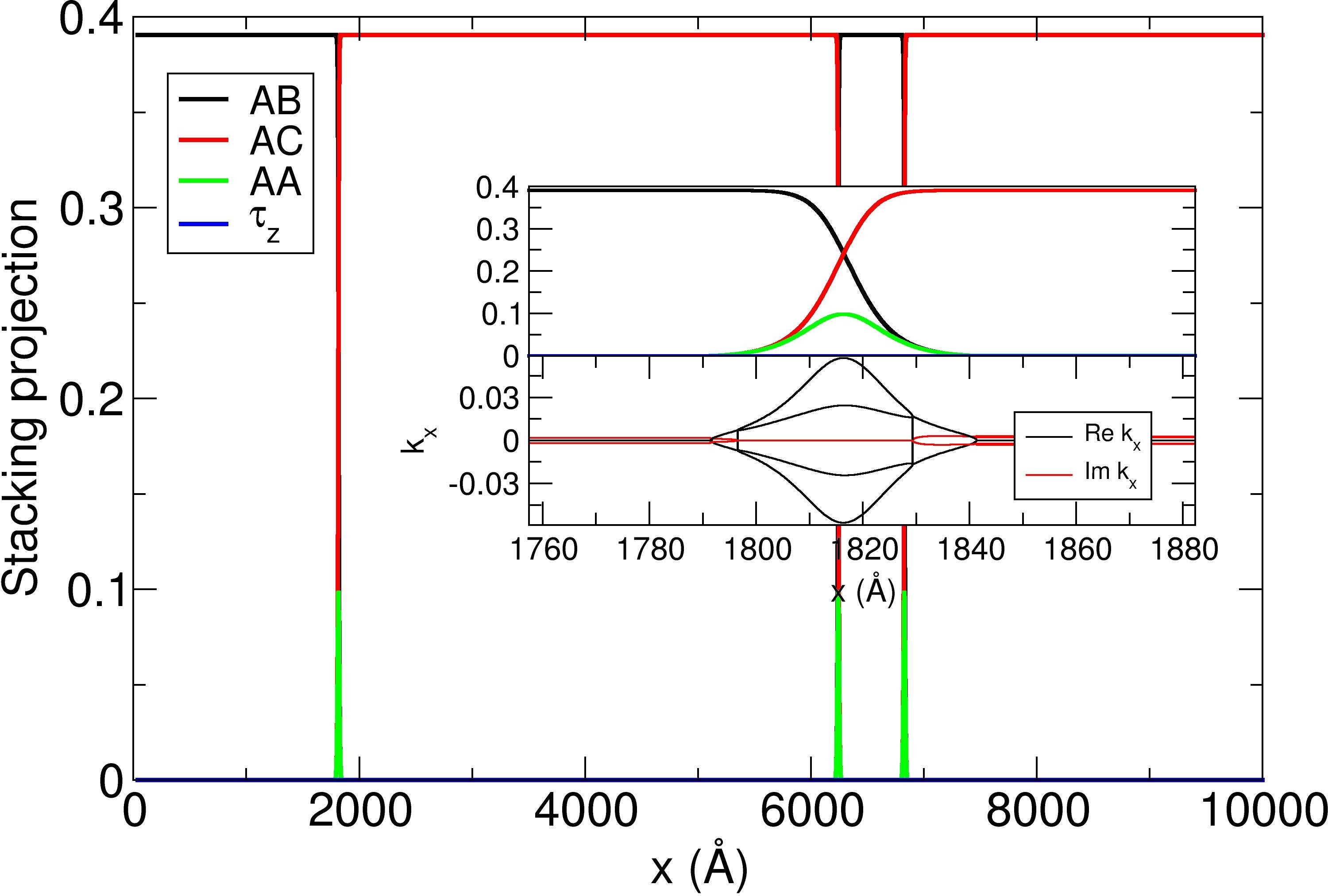}
    \caption{Effective interlayer stacking field for a partial dislocation array of 3 type 3 partials (the situation in which the partial Burgers vector $\bd_3$ is perpendicular to the transport direction) positioned in a device of 1$\mu$m length. The field is projected onto the complete set of stacking matrices defined in Eq.~\ref{stack} of the text that indicate, as shown in the caption, the AB, AC, AA, and $\tau_z$ components of the stacking field. The upper inset panel displays a close up of the first partial, and the lower inset panel the $k_x$ momenta plotted over the same region of space. This partial geometry is that for which the wavefunction is calculated in Fig.~2.
}
    \label{pda}
\end{figure}

\emph{Electronic structure}: The key ingredient in describing, in the effective Hamiltonian approach, the electronic structure of a system with a complex spatially dependent interlayer stacking is an interlayer effective field that encodes an arbitrary interlayer structural deformation\cite{Kisslinger,M}:

\begin{equation}
\label{M}
S(\br) = \sum_{i} M_i t_\perp(K_i) e^{-i\bK_i.\Delta \bu(\br)}
\end{equation}
where $\Delta \bu(\br)$ is a field that describes the interlayer deformation in terms of a local shift of the two layers, $t_\perp(q)$ the Fourier transform of the interlayer hopping function, and $\left[M_i\right]_{\alpha \beta} = e^{i\bG_i.(\bnu_\alpha-\bnu_\beta)}$ with $\{\bK_i\}$ the translation group of the high symmetry $K$ point $\bK_1 = (2/3,0)$, $\{\bG_i\}$ the corresponding reciprocal lattice vectors, and $\bnu_\alpha$ ($\bnu_\beta$) the basis vectors of layer one (two) of the bilayer. This evidently treats AB, AC, and any intermediate stacking, on the same footing (the standard AB and AC interlayer fields may, for example, be recovered by setting $\Delta\bu = {\bf 0}$ and $\Delta \bu = \bd_i$, the nearest neighbour vectors of graphene), and therefore provides a consistent framework for describing samples threaded by a partial dislocation network in which all stacking types are present. The $S(\br)$ interlayer field is best represented by a projection onto 4 stacking matrices which, with a $c$-number coefficient, provides for the 8 unknowns of the layer off-diagonal blocks: $S_0(\br) = c_{AB} \tau_{AB} + c_{AC} \tau_{AC} + c_{AA} \tau_{AA} + c_z \tau_z$. These matrices are given by

\begin{equation}
\begin{array}{cccc}
 \tau_{AB} = \begin{pmatrix} 1 & 0 \\ 0 & 0 \end{pmatrix}, &
 \tau_{AC} = \begin{pmatrix} 0 & 0 \\ 0 & 1 \end{pmatrix}, &
 \tau_{AA} = \begin{pmatrix} 0 & 1 \\ 1 & 0 \end{pmatrix}, &
 \tau_z = \begin{pmatrix} 0 & 1 \\ -1 & 0 \end{pmatrix}
\end{array}
\label{stack}
\end{equation}
and this projection is therefore directly informative of the local stacking of the bilayer. In Fig.~\ref{pda} we show the interlayer effective field for an example partial dislocation array, along with a zoom of one of the partial dislocations. Also shown in the inset is the momentum in the transport direction, $k_x$, which acquires a real component within the dislocation, but is pure imaginary within the AB and AC domains.

In addition to this interlayer ``stacking field'' a partial dislocation necessarily involves strain. This we model based on structural calculations\cite{Butz}, and implement following the standard effective field theory approach by layer diagonal scalar $V\!\propto\! u^{(n)}_{xx}+u^{(n)}_{yy}$ and pseudo-gauge $\bA \!\propto\! (u^{(n)}_{yy}-u^{(n)}_{xx},2u^{(n)}_{xy})$ fields in the Hamiltonian ($u^{(n)}_{ij}$ is the deformation tensor of layer $n$).

\emph{Transport calculation}: For the transport calculation we use the Landauer formalism, especially appropriate as we are (a) interested in energies close to or at the Dirac point and (b) the experimental system is ultra-clean. Diffusive transport (which would mandate the use of the Boltzmann or Kubo formalism) is therefore not expected to be a physically relevant description. The system scattering matrix we obtain via the transfer matrix scheme implemented by a standard decimation technique in which the 1$\mu$m sample is represented by a series of strips of small width (typically, $100\,$\AA~for the terraces and $0.2\,$\AA~for the partial dislocation regions). The dispersion relation for bilayer graphene

\vspace{-0.4cm}
\begin{equation}
k_x = \pm \left[\frac{E(E\pm t_\perp)}{(\hbar v_F)^2} - k_y^2\right]^{1/2},
\label{bspec}
\end{equation}
implies that for finite lead momenta $k_y$ and $E\approx 0$ the propagation momenta $k_x$ is imaginary, a fact that plays a crucial role in the existence of the minimal metallic conductivity in graphene and bilayer graphene~\cite{Twor,Snyman}. Propagating the transfer matrix scheme in such a situation, however, is problematic. The presence of both exponentially positive and exponentially negative terms in the transfer matrices leads rapidly to a loss of numerical accuracy upon multiplication. Fortunately, a solution is well known in the context of propagating Maxwell's equations in photonic materials~\cite{Pendry}; one uses the scattering matrix instead of the transfer matrix for propagation. This approach may straightforwardly be adapted to the case of graphene physics.

\section{Author contributions}

The project was framed by HW and SS. The code development and calculations were performed by SS. All authors wrote the manuscript and participated in the analysis of the results.

\section{Additional information}
\subsection{Competing financial interests}

The authors declare no competing financial interests.

\subsection{Acknowledgments}

The work was carried out in the framework of the SFB 953 and SPP 1243 of the Deutsche Forschungsgemeinschaft (DFG).


\end{document}